\documentstyle[12pt,version2,aps,preprint]{revtex}

\def\sbox#1{\mbox{\scriptsize #1}}
\def\ch{{\mbox{ch}}}
\def\sh{{\mbox{sh}}}
\def\v#1{\mbox{\boldmath$#1$}}
\def\vs#1{\mbox{\small \boldmath$#1$}}

\def\simgeq{\mbox{\raisebox{-1.0ex}{$\stackrel{>}{\sim}$}}}
\begin{document}
\begin{title}
{\bf Modified Spin Wave Thoery \\ of the Bilayer Square Lattice \\ Frustrated
Quantum Heisenberg Antiferromagnet
}
\end{title}
\author{
 Kazuo HIDA}

\begin{instit}
\it
Department of Physics, Faculty of Science, \\ Saitama University, Urawa,
Saitama 338
\end{instit}

\begin{center}
(Received\hspace{4cm})
\end{center}

The ground state of the square lattice bilayer quantum antiferromagnet with
nearest and next-nearest neighbour intralayer interaction is studied by means
of the modified spin wave method. For weak interlayer coupling, the ground
state is found to be always magnetically ordered while the quantum disordered
phase appear for large enough interlayer coupling. The properties of the
disordered phase vary according to the strength of the frustration. In the
regime of weak frustration, the disordered ground state is an almost
uncorrelated assembly of interlayer dimers, while in the strongly frustrated
regime the quantum spin liquid phase which has considerable N\'eel type short
range order appears. The behavior of the sublattice magnetization and spin-spin
correlation length in each phase is discussed.

\vspace{2mm}

\noindent
e-mail: hida@th.phy.saitama-u.ac.jp

\newpage

\section{Introduction}

The spin-1/2 square lattice Heisenberg model is now widely believed to have an
antiferromagnetic long range order in the ground
state.\cite{chn1,ry1,sgh1,kk1,kls1} However, it is expected that the strong
quantum fluctuation in this system may lead to the destruction of the long
range order with the help of some additional mechanism.
In this context, the square lattice antiferromagnetic Heisenberg model with
nearest and next-nearest exchange interaction (hereafter called $J_1-J_2$
model)\cite{cd1,ns1,mpb1,sdt1,nh1,ir1,ok1,gsh1,jf1,efj1,igg1,szp1,es1} and the
bilayer Heisenberg model\cite{mh1,kh1,kh2,mm1,ss1} have been studied
extensively. Both of these models are expected to have the quantum disordered
ground state for appropriate parameter regime. However, because the mechanisms
leading to the quantum disordered phase in these two models are of very
different nature, it must be most interesting to study their interplay in the
bilayer $J_1-J_2$ model.

In the $J_1-J_2$ model, the competition between the nearest neighbour
interaction $J_1$ and the nearest neighbour interaction $J_2$ introduces the
frustration in spin configuration which enhances the quantum fluctuation.
However, the conclusion about the presence of the quantum disordered state in
this model is still controvertial even in the most frustrated regime.

On the other hand, in the bilayer model, if the interlayer antiferromagnetic
coupling is strong enough, the spins on both layers form interlayer singlet
pairs and the quantum fluctuation is enhanced leading to the quantum disordered
state.
In other words, the antiferromagnetic interlayer coupling reduces the effective
spin magnitude. Therefore this model may be regarded as the physical
realization of the single layer Heisenberg model with spin {\it less than} 1/2.
Actually, in the study of the single layer $J_1-J_2$ model, there are
considerable number of works which cast doubt on the presence of quantum
disordered phase even for $S=1/2$ and $J_2/J_1 = 0.5$.\cite{ns1,mpb1,sdt1,nh1}
But some of these works also predict the presence of quantum disordered phase
for $S < 1/2$ which is unreachable within the single layer
model.\cite{ns1,mpb1} The bilayer$J_1-J_2$ model can effectively realize such
situation.

This paper is organized as follows: The bilayer $J_1-J_2$ model and its
clasical ground state are explained in the next section. In section 3, the
modified spin wave approximation\cite{mt1} is applied to this model.  The phase
diagram and the behavior of physical quantities are presented in section 4. The
last section is devoted to summary and discussion.

\section{Bilayer $J_1-J_2$ Model}

The Hamiltonian of the bilayer $J_1-J_2$ model is given as follows,

\begin{equation}
\label{eq:ham1}
H =J_1\sum_{<i,j>_{nn}}(\v{S}^A_{i} \v{S}^A_{j}+\v{S}^B_{i} \v{S}^B_{j})
+ J_2\sum_{<i,j>_{nnn}}(\v{S}^A_{i} \v{S}^A_{j}+\v{S}^B_{i} \v{S}^B_{j})
+ J_3\sum_i \v{S}^A_{i} \v{S}^B_{i},
\end{equation}
where $\v{S^{\mu}_i}$ is the spin operator with magnitude $S$ on the $i$-th
site of the layer $\alpha (\mu = A \mbox{ or } B )$. The expression
$\displaystyle\sum_{<i,j>_{nn}}$ and $\displaystyle\sum_{<i,j>_{nnn}}$ denote
the summation over the intralayer nearest neighbour pairs and next nearest
neighbour pairs, respectively. The last term represents the interlayer
coupling. All exchange couplings are assumed to be antiferromagnetic. In the
following, we denote the ratios $J_2/J_1 = \alpha$ and $J_3/J_1 = \beta$ and
take the energy unit $J_1=1$.

In the classical limit, the ground state is the N\'eel state and the collinear
state according as $\alpha <0.5$ or $\alpha > 0.5$. Actually, if the quantum
fluctuation is completely neglected, infinite number of ground state
configurations are degenerate for $\alpha > 0.5$.\cite{cd1,kk2} However, for
the single layer $J_1-J_2$ model, it is known that this degeneracy is lifted by
the quantum fluctuation and the collinear ground state is chosen.\cite{kk2} It
is straightforward to extend this argument to the bilayer model. Therefore, in
the following, we only consider the collinear order for $\alpha > 0.5$.

Thus we consider the following two types of spin configuration in the classical
limit;

\begin{equation}
\label{eq:cln}
%\begin{eqnarray}
\left\{\begin{array}{ll}
S_i^{Az}=S(-1)^{m+n} \\
S_i^{Bz}=S(-1)^{m+n+1}
\end{array}\right.
\mbox{: N\'eel (N-)configuration,} \\
\end{equation}
\begin{equation}
\label{eq:clc}
\left\{\begin{array}{ll}
S_i^{Az}=S(-1)^{m} \\
S_i^{Bz}=S(-1)^{m+1}
\end{array}\right.
\mbox{: collinear (C-)configuration,}
%\end{eqnarray}
\end{equation}
where the position of the $i$-th site $\v{r}_i$ is denoted by $(m,n)$. In the
following, we treat the quantum fluctuations around these configurations by
means of the modified spin wave method.\cite{mt1}

\section{Modified spin wave approximation}

Based on the classical configurations explained in the last section, let us
introduce the Dyson-Maleev transformation\cite{fjd1,svm1} for each spin as
follows:

\begin{eqnarray}
&&S_i^{Az}=(-1)^{m+n}(S-a_i^{\dagger}a_i),  \\
&&S_i^{Bz}=(-1)^{m+n+1}(S-b_i^{\dagger}b_i), \\
&&\left.\begin{array}{ll}
S_i^{A+}=\sqrt{\displaystyle\frac{1}{2S}}(1-a_i^{\dagger}a_i)a_i, &
S_i^{B+}=-b_i^{\dagger}\sqrt{\displaystyle\frac{1}{2S}}(1-b_i^{\dagger}b_i)\\
S_i^{A-}=\sqrt{2S}a_i^{\dagger}, & S_i^{B-}=-\sqrt{2S}b_i \\
\end{array}
\right\}\mbox{for}\ \ m+n=\mbox{even,} \\
&&\left.\begin{array}{ll}
S_i^{A+}=-a_i^{\dagger}\sqrt{\displaystyle\frac{1}{2S}}(1-a_i^{\dagger}a_i), &
S_i^{B+}=\sqrt{\displaystyle\frac{1}{2S}}(1-b_i^{\dagger}b_i)b_i \\
S_i^{A-}=-\sqrt{2S}a_i, & S_i^{B-}=\sqrt{2S}b_i^{\dagger} \\
\end{array}
\right\}\mbox{for}\ \ m+n=\mbox{odd}, \\
\end{eqnarray}
for the N-configuration (\ref{eq:cln}), and

\begin{eqnarray}
&&S_i^{Az}=(-1)^{m}(S-a_i^{\dagger}a_i),  \\
&&S_i^{Bz}=(-1)^{m+1}(S-b_i^{\dagger}b_i), \\
&&\left.\begin{array}{ll}
S_i^{A+}=\sqrt{\displaystyle\frac{1}{2S}}(1-a_i^{\dagger}a_i)a_i, &
S_i^{B+}=-b_i^{\dagger}\sqrt{\displaystyle\frac{1}{2S}}(1-b_i^{\dagger}b_i) \\
S_i^{A-}=\sqrt{2S}a_i^{\dagger}, & S_i^{B-}=-\sqrt{2S}b_i \\
\end{array}
\right\}\mbox{for}\ \ m=\mbox{even}, \\
&&\left.\begin{array}{ll}
S_i^{A+}=-a_i^{\dagger}\sqrt{\displaystyle\frac{1}{2S}}(1-a_i^{\dagger}a_i), &
S_i^{B+}=\sqrt{\displaystyle\frac{1}{2S}}(1-b_i^{\dagger}b_i)b_i \\
S_i^{A-}=-\sqrt{2S}a_i, & S_i^{B-}=\sqrt{2S}b_i^{\dagger} \\
\end{array}
\right\}\mbox{for}\ \ m=\mbox{odd}, \\
\end{eqnarray}
for the C-configuration (\ref{eq:clc}).

For the N-configuration (\ref{eq:cln}), the Hamiltonian (\ref{eq:ham1}) is
rewritten as,
\begin{eqnarray}
\label{eq:hamb1}
H &=&\sum_{<i,j>_{nn}}\{-2S^2+S(a^{\dagger}_i a_i+a^{\dagger}_j a_j -
a^{\dagger}_ia^{\dagger}_j  - a_ia_j  \nonumber \\
&+& b^{\dagger}_i b_i+b^{\dagger}_j b_j - b^{\dagger}_i b^{\dagger}_j  - b_i
b_j) + a^{\dagger}_i(a^{\dagger}_j - a_i)^2a_j/2
 + b^{\dagger}_i(b^{\dagger}_j - b_i)^2 b_j/2\} \nonumber \\
&+&\alpha \sum_{<i,j>_{nnn}}\{2S^2 - S(a^{\dagger}_i -a^{\dagger}_j)(a_i - a_j)
- a^{\dagger}_ia^{\dagger}_j(a_i-a_j)^2\delta_{ie}/2
- (a^{\dagger}_i-a^{\dagger}_j)^2a_ia_j \delta_{io}/2  \nonumber \\
&-& S(b^{\dagger}_i -b^{\dagger}_j)(b_i - b_j) -
b^{\dagger}_ib^{\dagger}_j(b_i-b_j)^2\delta_{io}/2
-(b^{\dagger}_i-b^{\dagger}_j)^2 b_ib_j\delta_{ie}/2\} \nonumber \\
&+&\beta
\sum_i\{-S^2+S(a^{\dagger}_i a_i+b^{\dagger}_i b_i - a^{\dagger}_i
b^{\dagger}_i  - a_i b_i) + a^{\dagger}_i(b^{\dagger}_i - a_i)^2b_i/2  \},
\end{eqnarray}
where $\delta_{ie}(\delta_{io}) = 1$ or 0 according as $m+n =$ even (odd) or
odd(even) where $\v{r}_i=(m,n)$. For the C-configuration (\ref{eq:clc}), we
have
\begin{eqnarray}
\label{eq:hamb2}
H &=& \sum_{<i,j>_{nnx}}\{-2S^2+S(a^{\dagger}_i a_i+a^{\dagger}_j a_j -
a^{\dagger}_i a^{\dagger}_j  - a_i a_j  \nonumber \\
&+& b^{\dagger}_i b_i+b^{\dagger}_j b_j - b^{\dagger}_i b^{\dagger}_j  - b_i
b_j)+ a^{\dagger}_i(a^{\dagger}_j - a_i)^2a_j/2
 + b^{\dagger}_i(b^{\dagger}_j - b_i)^2 b_j/2\} \nonumber \\
&+& \sum_{<i,j>_{nny}}\{2S^2 - S(a^{\dagger}_i -a^{\dagger}_j)(a_i - a_j) -
a^{\dagger}_ia^{\dagger}_j(a_i-a_j)^2 \delta_{iey}/2
- (a^{\dagger}_i-a^{\dagger}_j)^2 a_ia_j\delta_{ioy}/2 \nonumber \\
&-&S(b^{\dagger}_i -b^{\dagger}_j)(b_i - b_j) -(b^{\dagger}_i-b^{\dagger}_j)^2
b_ib_j \delta_{iey}/2
 - b^{\dagger}_ib^{\dagger}_j(b_i-b_j)^2\delta_{ioy}/2\} \nonumber \\
 &+&\alpha \sum_{<i,j>_{nnn}}\{-2S^2+S(a^{\dagger}_i a_i+a^{\dagger}_j a_j -
a^{\dagger}_i a^{\dagger}_j  - a_i a_j  \nonumber \\
&+& b^{\dagger}_i b_i+b^{\dagger}_j b_j - b^{\dagger}_i b^{\dagger}_j  - b_i
b_j) + a^{\dagger}_i(a^{\dagger}_j - a_i)^2a_j/2
 + b^{\dagger}_i(b^{\dagger}_j - b_i)^2 b_j/2\} \nonumber \\
&+&\beta
\sum_i\{-S^2+S(a^{\dagger}_i a_i+b^{\dagger}_i b_i - a^{\dagger}_i
b^{\dagger}_i  - a_i b_i) + a^{\dagger}_i(b^{\dagger}_i - a_i)^2b_i/2  \},
\end{eqnarray}
where $\delta_{iey}(\delta_{ioy}) = 1$ or 0 according as $m =$ even (odd) or
odd(even). The notations  $\displaystyle\sum_{<i,j>_{nnx}}$ and
$\displaystyle\sum_{<i,j>_{nny}}$ denote the summation over the intralayer
nearest neighbour pairs in $x$ and $y$ direction, respectively.

Following Takahashi,\cite{mt1}  we assume the constraint that the sublattice
magnetization vanish as expected for the two dimensional spin system with
continuous symmetry at finite temperatures.\cite{mw1}
\begin{equation}
\label{eq:const}
S = \sum_i<a_i^{\dagger}a_i>= \sum_i<b_i^{\dagger}b_i>.
\end{equation}
We impose this condition even in the ground state where the long range
sublattice magnetization may be present. This means that the average is taken
over the direction of the sublattice magnetization even in the ordered phase.
Nevertheless, we can calculate the sublattice magnetization from the long range
part of the correlation function which originate from the Bose condensate of
the bose fields $a_i$ and $b_i$.\cite{kh1,mt1} Although the validity of this
procedure is not well founded, these are the common features of the modified
spin wave method and we do not discuss this point further.

We treat the nonlinear terms in (\ref{eq:hamb1}) and (\ref{eq:hamb2}) by the
mean field approximation as.

\begin{eqnarray}
\label{eq:hmfn}
H^{\sbox{MF}} &=&\sum_{<i,j>_{nn}}\{\Delta(a^{\dagger}_i a_i+a^{\dagger}_j a_j
- a^{\dagger}_i a^{\dagger}_j  - a_i a_j
+ b^{\dagger}_i b_i+b^{\dagger}_j b_j - b^{\dagger}_i b^{\dagger}_j  - b_i b_j)
+ 2\Delta^2 - 4S\Delta\} \nonumber \\
&+&\alpha \sum_{<i,j>_{nnn}}\{-q_{xy}(a^{\dagger}_i -a^{\dagger}_j)(a_i - a_j)
- q_{xy}(b^{\dagger}_i -b^{\dagger}_j)(b_i - b_j)+4Sq_{xy}-2q_{xy}^2\}
\nonumber \\
&+&\beta
\sum_i\{\Delta_{\sbox{AB}}(a^{\dagger}_i a_i+b^{\dagger}_i b_i - a^{\dagger}_i
b^{\dagger}_i- a_i b_i) +\Delta_{\sbox{AB}}^{2}-2S\Delta_{\sbox{AB}}\}
\nonumber \\
&-&\sum_i \{\mu (S - a_i^{\dagger}a_i)+ \mu (S - b_i^{\dagger}b_i)\},
\end{eqnarray}
for the N-configuration phase. Here $\mu$ is the Lagrangian multiplier
corrresponding to the constraint (\ref{eq:const}). The order parameters are
defined by
\begin{eqnarray}
&&<a^{\dagger}_ia_j> = <b^{\dagger}_ib_j> = q_{xy}\ \  \mbox{for }\
\v{r}_j=\v{r_i} + \v{\delta},
 \\
&&<a^{\dagger}_ia^{\dagger}_j> =<b^{\dagger}_ib^{\dagger}_j> = \Delta \ \
\mbox{for}\ \  \v{r}_j=\v{r_i} + \v{\rho}, \\
&&<a^{\dagger}_ib^{\dagger}_i> =<a_ib_i> = \Delta_{\sbox{AB}},
\end{eqnarray}
where $\v{\rho}$ is the vector to the nearest neighbour sites and $\v{\delta}$
to the next nearest sites.

For the C-configuration, we have
\begin{eqnarray}
\label{eq:hmfc}
H^{\sbox{MF}} &=& \sum_{<i,j>_{nnx}}\{\Delta(a^{\dagger}_i a_i+a^{\dagger}_j
a_j - a^{\dagger}_i a^{\dagger}_j  - a_i a_j
+ b^{\dagger}_i b_i+b^{\dagger}_j b_j - b^{\dagger}_i b^{\dagger}_j  - b_i b_j)
+ 2\Delta^2 - 4S\Delta\} \nonumber \\
&+& \sum_{<i,j>_{nny}}\{-q_y(a^{\dagger}_i -a^{\dagger}_j)(a_i - a_j)
- q_y(b^{\dagger}_i -b^{\dagger}_j)(b_i - b_j)+4Sq_y-2q_{y}^2\}  \nonumber \\
 &+&\alpha \sum_{<i,j>_{nnn}}\{\Delta_{xy}(a^{\dagger}_i a_i+a^{\dagger}_j a_j
- a^{\dagger}_i a^{\dagger}_j  - a_i a_j
+ b^{\dagger}_i b_i+b^{\dagger}_j b_j - b^{\dagger}_i b^{\dagger}_j  - b_i b_j)
+ 2\Delta_{xy}^2 - 4S\Delta_{xy}\} \nonumber \\
&+&\beta
\sum_i\{\Delta_{\sbox{AB}}(a^{\dagger}_i a_i+b^{\dagger}_i b_i - a^{\dagger}_i
b^{\dagger}_i- a_i b_i) +\Delta_{\sbox{AB}}^{2}-2S\Delta_{\sbox{AB}}\}
\nonumber \\
&-&\sum_i \{\mu (S - a_i^{\dagger}a_i)+ \mu (S - b_i^{\dagger}b_i)\}.
\end{eqnarray}
Here, the order parameters are defined by
\begin{eqnarray}
&&<a^{\dagger}_ia_j> = <b^{\dagger}_ib_j> = q_y \ \  \mbox{for }\
\v{r}_j=\v{r_i} + \v{\rho_y},
 \\
&&<a^{\dagger}_ia^{\dagger}_j> =<b^{\dagger}_ib^{\dagger}_j> = \Delta_x\ \
\mbox{for}\ \  \v{r}_j=\v{r_i} + \v{\rho_x}, \\
&&<a^{\dagger}_ia^{\dagger}_j> =<b^{\dagger}_ib^{\dagger}_j> = \Delta_{xy}\ \
\mbox{for}\ \  \v{r}_j=\v{r_i} + \v{\delta} ,\\
&&<a^{\dagger}_ib^{\dagger}_i> =<a_ib_i> = \Delta_{\sbox{AB}},
\end{eqnarray}
where $\v{\rho}_x$ and $\v{\rho}_y$ is the vector to the nearest neighbour
sites in $x$ and $y$ direction, respectively.

These mean field Hamiltonians are transformed into the fourier space as

\begin{eqnarray}
\label{eq:mfham}
H^{\sbox{MF}} &=& \sum_{\vs{k}}{}^>\{ \Lambda(\v{k})(a^{\dagger}_{\vs{k}}
a_{\vs{k}} + a^{\dagger}_{-\vs{k}} a_{-\vs{k}}+ b^{\dagger}_{\vs{k}} b_{\vs{k}}
+ b^{\dagger}_{-\vs{k}} b_{-\vs{k}}) \\ \nonumber
&-& \Gamma(\v{k})(a^{\dagger}_{\vs{k}} a^{\dagger}_{-\vs{k}} + a_{\vs{k}}
a_{-\vs{k}} + b^{\dagger}_{\vs{k}} b^{\dagger}_{-\vs{k}}+ b_{\vs{k}}
b_{-\vs{k}} ) \\ \nonumber
&+& \beta \Delta_{\sbox{AB}}(a^{\dagger}_{\vs{k}} a_{\vs{k}} +
a^{\dagger}_{-\vs{k}} a_{-\vs{k}}+ b^{\dagger}_{\vs{k}} b_{\vs{k}} +
b^{\dagger}_{-\vs{k}} b_{-\vs{k}} - a^{\dagger}_{\vs{k}} b^{\dagger}_{-\vs{k}}
- a_{\vs{k}} b_{-\vs{k}} - b^{\dagger}_{\vs{k}} a^{\dagger}_{-\vs{k}}-
b_{\vs{k}} a_{-\vs{k}} )\} \nonumber \\
&+&E_0,
\end{eqnarray}
where
\begin{eqnarray}
\Lambda(\v{k}) &=& 4(\Delta - \alpha q_{xy} (1-\cos(k_x)\cos(k_y))) + \mu ,\\
\Gamma(\v{k})&=& 4\Delta\gamma(\v{k}),\\
\gamma(\v{k}) &=& \frac{1}{2}(\cos(k_x)+\cos(k_y)), \\
E_0 &=&N\{4\Delta^2 - 8S\Delta+4\alpha(4Sq_{xy}-2q_{xy}^2) \\
&+&\beta(\Delta_{\sbox{AB}}^{2}-2S\Delta_{\sbox{AB}})-2\mu S \},
\end{eqnarray}
for the N-configuration  and
\begin{eqnarray}
\Lambda(\v{k}) &=& -2 q_y(1-\cos(k_y)) + \mu, \\
\Gamma(\v{k}) &=& 2\Delta_x \cos(k_x) + 4\alpha \Delta_{xy}\cos(k_x)\cos(k_y),
\\
E_0 &=& N\{(2\Delta_x^2 - 4S\Delta_x + 4Sq_y-2q_{y}^2+2\alpha(2\Delta_{xy}^2 \\
&-& 4S\Delta_{xy})+\beta(\Delta_{\sbox{AB}}^{2}-2S\Delta_{\sbox{AB}})-2 \mu S
\},
\end{eqnarray}
for the C-configuration. The summation $\displaystyle{\sum_{\vs{k}}{}^>}$ is
taken over the left half of the Brillouin zone ($k_x > 0$) because the momenta
$\v{k}$ and $-\v{k}$ are explicitly wrtten in (\ref{eq:mfham}). The number of
the lattice sites in each layer is denoted by $N$.

These Hamiltonians are diagonalized as
\begin{equation}
H^{\sbox{MF}}= \sum_{\vs{k}} E_+(\v{k}) \alpha_{\vs{k}}^{\dagger}
\alpha_{\vs{k}} +  E_-(\v{k}) \beta_{\vs{k}}^{\dagger} \beta_{\vs{k}} +
E_{\sbox{G}},
\end{equation}
where
\begin{equation}
E_{\pm}(\v{k}) = \sqrt{\eta(\v{k})^2-(\Gamma(\v{k}) \pm \delta)^2},\ \ \delta
\equiv \beta\Delta_{\sbox{AB}},\ \ \eta(\v{k}) \equiv \Lambda(\v{k})+\delta,
\end{equation}
by the Bogoliubov transform
\begin{eqnarray}
a_{\vs{k}} =\{\ch \theta^{+}_{\vs{k}}\alpha_{\vs{k}} + \ch
\theta^{-}_{\vs{k}}\beta_{\vs{k}} +\sh
\theta^{+}_{\vs{k}}\alpha^{\dagger}_{-\vs{k}} + \sh
\theta^{-}_{-\vs{k}}\beta^{\dagger}_{-\vs{k}} \}/\sqrt{2}, \\
b_{\vs{k}} =\{\ch \theta^+_{\vs{k}}\alpha_{\vs{k}} - \ch
\theta^-_{\vs{k}}\beta_{\vs{k}} +\sh
\theta^+_{\vs{k}}\alpha^{\dagger}_{-\vs{k}} - \sh
\theta^-_{-\vs{k}}\beta^{\dagger}_{-\vs{k}} \}/\sqrt{2}, \\
a^{\dagger}_{-\vs{k}} =\{\sh \theta^+_{\vs{k}}\alpha_{\vs{k}} + \sh
\theta^-_{\vs{k}}\beta_{\vs{k}} +\ch
\theta^+_{\vs{k}}\alpha^{\dagger}_{-\vs{k}} + \ch
\theta^-_{-\vs{k}}\beta^{\dagger}_{-\vs{k}} \}/\sqrt{2}, \\
b^{\dagger}_{-\vs{k}} =\{\sh \theta^+_{\vs{k}}\alpha_{\vs{k}} - \sh
\theta^-_{\vs{k}}\beta_{\vs{k}} +\ch
\theta^+_{\vs{k}}\alpha^{\dagger}_{-\vs{k}} - \ch
\theta^-_{-\vs{k}}\beta^{\dagger}_{-\vs{k}} \}/\sqrt{2},
\end{eqnarray}
where
\begin{equation}
\ch \theta^{\pm}_{\vs{k}} =
\sqrt{\frac{1}{2}\left(\frac{\eta(\v{k})}{E_{\pm}(\v{k})}+1\right)}, \ \ \
\sh \theta^{\pm}_{\vs{k}} =
\sqrt{\frac{1}{2}\left(\frac{\eta(\v{k})}{E_{\pm}(\v{k})}-1\right)}
\mbox{sgn}(\Gamma_{\vs{k}}\pm \delta).
\end{equation}
The self consistent equations for the order parameters are
\begin{eqnarray}
\label{eq:scn1}
\Delta &=&  \frac{1}{N} \sum_{\v{k},\pm}{}'
\frac{\Gamma(\v{k})\pm\delta}{4E_{\pm}(\v{k})}\gamma(\v{k}) +N_0^{\sbox{N}} ,\\
\label{eq:scn2}
\Delta_{\sbox{AB}} &=&  \frac{1}{N} \sum_{\v{k}, \pm}{}'
\frac{\delta\pm\Gamma(\v{k})}{4E_{\pm}(\v{k})}+N_0^{\sbox{N}}, \\
\label{eq:scn3}
q_{xy} &=&  \frac{1}{N} \sum_{\v{k}, \pm}{}'
\frac{\eta(\v{k})}{4E_{\pm}(\v{k})}\cos(k_x)\cos(k_y)+N_0^{\sbox{N}},
\end{eqnarray}
for the N-configuration and
\begin{eqnarray}
\label{eq:scc1}
\Delta_{xy} &=&  \frac{1}{N} \sum_{\v{k}, \pm}{}'
\frac{\Gamma(\v{k})\pm\delta}{4E_{\pm}(\v{k})}\cos(k_x)\cos(k_y)
+N_0^{\sbox{C}}, \\
\label{eq:scc2}
\Delta_{x} &=&  \frac{1}{N} \sum_{\v{k}, \pm}{}'
\frac{\Gamma(\v{k})\pm\delta}{4E_{\pm}(\v{k})}\cos(k_x) +N_0^{\sbox{C}},  \\
\label{eq:scc3}
\Delta_{\sbox{AB}} &=&  \frac{1}{N} \sum_{\v{k}, \pm}{}'
\frac{\delta\pm\Gamma(\v{k})}{4E_{\pm}(\v{k})} +N_0^{\sbox{C}},  \\
q_y &=&  \frac{1}{N} \sum_{\v{k}, \pm}{}'
\frac{\eta(\v{k})}{4E_{\pm}(\v{k})}\cos(k_y)+N_0^{\sbox{C}},
\end{eqnarray}
for the C-configuration. The summation $\displaystyle{ \sum_{\v{k}, \pm}{}'}$
excludes $\v{k}$'s for which $E_{\pm}(\v{k})=0$. The quantities
$N_0^{\sbox{N}}$ and $N_0^{\sbox{C}}$ are  the amplitudes of the Bose
condensate of the bose particles represented by the operators $\alpha_{\vs{k}}$
and  $\beta_{\vs{k}}$ in the N\'eel phase and collinear phase, respectively.
They correspond to the magnitude of the N\'eel or collinear long range
order.\cite{ns1,kh2,mt1} These quantities vanish in the disordered phase.
The constraint (\ref{eq:const}) is rewritten in the form
\begin{equation}
S = -\frac{1}{2} + \frac{1}{N} \sum_{\v{k}, \pm}{}'
\frac{\eta(\v{k})}{4E_{\pm}(\v{k})} +N_0,
\end{equation}
in both cases where $N_0$ stands for $N_0^{\sbox{C}} $ or $N_0^{\sbox{C}} $. In
order that $N_0$ remains finite, the excitation spectrum must have  zero modes.
This requirement fixes the value of $\mu$ in the ordered phase. On the other
hand, for the solution corresponding to the disordered phase, the energy
spectrum has no zero mode and the value of $\mu$ must be fixed so that the
self-consistent equations are satisfied with $N_0 =0$.
 The ground state energy $E_{\sbox{G}}$ is given as follows:
\begin{eqnarray}
E_{\sbox{G}} &=& E_{\sbox{G}}^{\sbox{N}} \equiv N(-4 \Delta^2  +4 \alpha
q_{xy}^2- \beta \Delta_{\sbox{AB}}^2)\ \  \mbox{: N-configuration}, \\
E_{\sbox{G}} &=& E_{\sbox{G}}^{\sbox{C}} \equiv N(-2 \Delta_{x}^2 + 2 q_y^2 -
4\alpha\Delta_{xy}^2 -\beta\Delta_{\sbox{AB}}^{2})\ \  \mbox{:
C-configuration}.
\end{eqnarray}
If the self consistent equations have more than two solutions, we choose the
thermodynamically stabe solution comparing the ground state energy.

\section{Phase Diagram}

We have solved numerically the set of equations for $S=1/2$ and found 4 kinds
of phases:

i) N\'eel ordered phase : $\Delta, \Delta_{\sbox{AB}}, q_{xy}, N_0^{\sbox{N}} >
0$

ii) Collinear ordered phase : $\Delta_{xy}, \Delta_{x}, \Delta_{\sbox{AB}},
q_y, N_0^{\sbox{C}} > 0$

iii) Spin liquid phase with N\'eel-type correlation (NSL phase): $\Delta,
\Delta_{\sbox{AB}}, q_{xy} >0, N_0^{\sbox{N}} = 0$

iv) Interlayer dimer phase (ILD phase):  $\Delta=q_{xy}=\Delta_{xy}=\Delta_{x}=
q_y=N_0^{\sbox{C}}=N_0^{\sbox{N}}=0,  \Delta_{\sbox{AB}}=\frac{\sqrt{3}}{2},$

The spin fuild phase with collinear-type correlation ( $\Delta_{xy},
\Delta_{x}, \Delta_{\sbox{AB}}, q_y >0 , N_0^{\sbox{C}} = 0$ ) is only found as
a metastable state. The ground state phase diagram is shown in Fig. \ref{fig1}.

For small $\beta$, there appears no disordered phase. The ground state changes
from the N\'eel ordered phase to the collinear ordered phase by the first order
transition around $\alpha \simgeq 0.5$. For larger values of $\beta$, however,
there appears a NSL phase around $\alpha \sim 0.5$. In this phase, the short
range intralayer singlet order parameter $\Delta$ remains finite, although the
long range order is absent. The transition between the N\'eel phase and NSL
phase is of the second order, while between the collinear ordered phase and NSL
phase the transition is of the first order. As the value of $\beta$ is further
increased, the intralayer correlation becomes weaker and the second order
transition to the ILD phase takes place at $\beta=4$. This is verified
analytically  from the stabilty analysis of Eqs.
(\ref{eq:scn1},\ref{eq:scn2},\ref{eq:scn3}) around the ILD solution. Of course,
this phase transition between the NSL phase and ILD phase is an artifact of the
mean field approximation and it should be interpreted as a crossover. For very
small or large value of $\alpha$, the direct first order transition from the
N\'eel or collinear ordered phase to the ILD phase takes place as in the
unfrustrated case.\cite{kh1}

It should be remarked that the spin-fluid phase with considerable intralayer
correlation is stabilized around $\alpha \sim 0.5$ where the effect of the
frustration is most pronounced. In this sense, our NSL state is the first
promising example of the non-trivial frustration induced quantum spin fluid
phase in two dimensions.

Figures \ref{fig2}(a) and \ref{fig2}(b) show the amplitude of the sublattice
magnetization $N_0^{\sbox{N}}$ and $N_0^{\sbox{C}}$ along the line with fixed
$\alpha$. In general, the sublattice magnetization is enhanced for small
$\beta$ and turns to decrease for larger $\beta$. Namely, the interlayer
coupling strengthens the ordering as far as it is small. The same is true also
in the unfrustrated case.\cite{kh1,kh2} This feature can be already observed
within the linear spin wave analysis.\cite{ooiwa}  Dotsenko\cite{ds1} has also
obtained the similar result using the mapping onto the nonlinear
$\sigma$-model.

The energy gap in the NLS phase $\Delta E$ is given by
\begin{equation}
\Delta E = E_+(\v{k}=0)=\sqrt{(8\Delta+\mu+\delta)(\mu-\delta)}.
\end{equation}
Figure \ref{fig3} shows the variation of $\Delta E$ for various values of
$\alpha$. It grows from 0 starting from the N\'eel-NSL boundary and saturates
at the NSL-ILD boundary. At the collinear-NSL boundary the energy gap in the
NSL phase remains finite reflecting the fact that this transition is of the
first order. However, we may expect that this is also an artifact of the
present approximation. Taking into account that $\Delta E$ is proportional to
the inverse of the spin-spin correlation length $\xi$ for $\xi >> 1$, the
antiferromagnetic short range correlation is highly enhanced in the NSL phase
near the phase boundary.

\section{Summary and Discussion}

The spin-1/2 bilayer $J_1-J_2$ model is studied by means of the modified spin
wave approximation and the ground state phase diagram is obtained. For small
interlayer coupling, N\'eel or collinear type long range order exists for any
value of $\alpha$. However, with the increase of the interlayer
antiferromagnetic coupling $\beta$, the correlated quantum spin fluid phase
appears around $\alpha \sim 0.5$. The width of the spin fluid phase becomes
wider as $\beta$ increases. On the other hand, the correlation length decreases
with the increase of $\beta$ and we only find the interlayer dimer phase for
$\beta > 4$.

It has been widely expected that the frustation effect enhance the quantum
flucutation and leads to the quantum spin liquid phase in two dimensional
system. Although the single layer $J_1-J_2$ model is one of such candidates,
the conclusion is rather sensitive to the approximations used, the method of
numerical calculations and data analysis. On the other hand, we may expect the
presence of the frustration induced highly correlated quantum spin liquid over
a wide range of parameters for the bilayer $J_1-J_2$ model.

We have also found that the N\'eel state remains stable for the value of
$\alpha$ slightly larger than 0.5. This may be explained as follows: In the
collinear phase, the classical ground state is continuously degenerate and
therefore the quantum fluctuation is more pronounced than the N\'eel phase.
Therefore the N\'eel state is stabilized rather than the collinear phase even
for $\alpha \simgeq 0.5$.

Using the modified spin wave approximation, Nishimori and Saika\cite{ns1}
obtained the result that the ground state energy jumps at the transition from
the N\'eel phase to the collinear phase in the spin-1/2 single layer $J_1-J_2$
model, while in our calculation this transition is a usual first order
transition even for $\beta =0$.
This is due to the fact that Nishimori and Saika expanded the ground state
energy with respect to $1/S$ and trunciated at the second order. Actually, such
estimation is known to give the results better than the mean field type
estimation in some cases.\cite{nm1} However, here we employ the na\"\i ve mean
field ground state energy without expansion for the consistency within the
present approximation. In any case, such details of the transition are beyond
the scope of our approximation.

Althouth we expect that our approach captures the essential features of the
ground state of the present model, our approximation is far from quantitative.
Even in the unfrustrated case ($\alpha = 0$), modified spin wave theory
predicts rather large N\'eel-ILS critical value of $\beta = \beta_c \sim
4.25$\cite{kh1} compared to the more reliable estimation $\beta_c \sim 2.5$ by
the dimer expansion\cite{kh2} and the quantum Monte Carlo simulation.\cite{ss1}
Unfortunately, the quantum Monte Carlo simulation is not expected to be
powerful enough for the frustrated model due to the negative sign problem. The
dimer expansion method may be promising but higher order calculation requires
excessive computational time and memory. Further investigation is thus required
for the quantitative understanding of the present model.

The author thanks K. Ooiwa for his collaboration in the early stage of this
work. This work is supported by the Grant-in-Aid for Scientific Research from
the Ministry of Education, Science and Culture. The numerical calculation is
performed by the HITAC S820/15 at the Information Processing Center of Saitama
University.

\newpage
\figure{The ground state phase diagram of the bilayer $J_1-J_2$ model. The
solid and broken lines are the lines of the first and second order transition,
respectively.
\label{fig1}}
\figure{The $\beta$-dependence of the sublattice magnetization in the (a)
N\'eel and (b) collinear ordered phases. The values of $\alpha$ are indicated
in the figure. The values on the phase boundary are represented by the open
circles. \label{fig2}}

\figure{The $\beta$-dependence of the excitation gap in the disordered phase.
The values of $\alpha$ are indicated in the figure. The values on the phase
boundary are represented by the open circles.
\label{fig3}}
\end{document}